\documentclass[sigconf]{acmart}

\copyrightyear{2023}
\acmYear{2023}
\setcopyright{acmlicensed}\acmConference[WWW '23]{Proceedings of the ACM Web Conference 2023}{May 1--5, 2023}{Austin, TX, USA}
\acmBooktitle{Proceedings of the ACM Web Conference 2023 (WWW '23), May 1--5, 2023, Austin, TX, USA}
\acmPrice{15.00}
\acmDOI{10.1145/3543507.3583462}
\acmISBN{978-1-4503-9416-1/23/04}

\acmConference[The Web Conference '23]{The Web Conference}{April 30--May 4,
  2023}{Austin, TX}

\usepackage{subcaption}
\usepackage{placeins}
\usepackage{caption}
\usepackage{comment}
\usepackage{multirow}

\usepackage{verbatim}
\usepackage{todonotes}
\presetkeys%
    {todonotes}%
    {inline,backgroundcolor=yellow}{}
\usepackage{xspace}
\usepackage{xcolor, colortbl}%

\begin{document}

\title{Cycle Generative Adversarial Networks for Complementary Item Recommendations}
\title{Complementary Item Recommendation with Cycle Generative Adversarial Networks}
\title{Semi-supervised Adversarial Learning for Complementary Item Recommendation}
\author{Koby Bibas}
\email{kobybibas@gmail.com}
\affiliation{%
  \institution{Meta}
  \city{Tel Aviv}
  \country{Israel}
}
\author{Oren Sar Shalom}
\email{oren.sarshalom@gmail.com}
\affiliation{%
  \institution{Amazon}
  \city{Tel Aviv}
  \country{Israel}
}
\authornote{Work was done while with Meta.}

\author{Dietmar Jannach}
\email{dietmar.jannach@aau.at}
\affiliation{%
  \institution{University of Klagenfurt}
  \city{Klagenfurt}
  \country{Austria}
}

\begin{abstract}
Complementary item recommendations are a ubiquitous feature of modern e-commerce sites. Such recommendations are highly effective when they are based on collaborative signals like co-purchase statistics. In certain online marketplaces, however, e.g., on online auction sites, constantly new items are added to the catalog. In such cases, complementary item recommendations are often based on item side-information due to a lack of interaction data. In this work, we propose a novel approach that can leverage both item side-information and
labeled complementary item pairs to generate effective complementary recommendations for cold items, i.e., for items for which no co-purchase statistics yet exist. Given that complementary items typically have to be of a different category than the seed item, we technically maintain a latent space for each item category. Simultaneously, we learn to project distributed item representations into these category spaces to determine suitable recommendations. The main learning process in our architecture utilizes labeled pairs of complementary items. In addition, we adopt ideas from Cycle Generative Adversarial Networks (CycleGAN) to leverage available item information even in case no labeled data exists for a given item and category. Experiments on three e-commerce datasets show that our method is highly effective. 
\end{abstract}


\maketitle

\section{Introduction}
\label{sec:introduction}
Many modern e-commerce sites provide \emph{complementary} item recommendations for their customers. For example, when online users shop for shirts, they may receive suggestions for suitable pairs of pants. In practice, such recommendations often appear under labels such as ``Frequently bought together'', and these recommendations are an important driver for cross-selling on e-commerce platforms~\cite{brovman2016optimizing,jannachjugovactmis2019,jannach2019towards}. Given the typically large product assortments in e-commerce, the ranking of complementary item recommendations is commonly based on behavioral patterns observed in the consumer community, for instance on co-purchase patterns. In addition, domain knowledge may be incorporated, for example, that the complementary item recommendations have to come from a pre-defined \emph{target} category (e.g., pants), or from a category that is different from the category of the \emph{seed} item, i.e., the item for which complementary suggestions are sought for.

In some domains, however, item catalogs can be very volatile~\cite{sar2015data}. 
On the eBay (\emph{ebay.com}) auction platform, for example, 
constantly new items become available and often there is only one exemplar of a certain product available~\cite{galron2018deep,wang2021personalized}.
Consequently, suitable methods are needed for \emph{cold} seed items, i.e., items for which no transaction data is available.
In this setting it is impossible to model the seed item based on collaborative signals like item co-purchases.
Thus, one may have to rely on alternative approaches, like processing side-information about the items, e.g., natural language descriptions of the items or structured item attributes.

Although we focus on cold items, it is still possible to leverage aggregate usage data. Specifically, such data allows us to infer a set of \emph{complementary categories}, i.e., categories that include complementary items, like the categories pants and shirts. Recommendations for an item from a certain category can then be constrained to items that belong to complementary categories. The advantages of this approach are threefold: (1) it helps rule out irrelevant recommendations; 
(2) it significantly reduces inference time; and (3) it can be used to design bundles \cite{zhu2014bundle, avny2022bruce}
(e.g., by recommending a single item per complementary category), and thereby allow users to understand how they were formed. Given these advantages, category-level constraints on complementary item recommendation is therefore a common practice at large e-commerce sites like
Facebook Shops and Instagram Shops\footnote{\url{https://research.facebook.com/blog/2021/12/shops-on-facebook-and-instagram-understanding-relationships-between-products-to-improve-buyer-and-seller-experience/}}. 



In this work, we propose a novel deep learning model to address this challenging problem setting.
The main technical idea of our approach is to maintain a latent space for each item category. Simultaneously, we learn to translate representations of the items, which we derive with the help of side-information, into these latent spaces. This process finally allows us to locate items in a given category space that are close to the seed item. The basic learning process in our architecture 
is based on \emph{supervised learning} using labeled pairs of complementary items. We note that these pairs of items, which can for example be obtained by analyzing existing shopping baskets of a site, do \emph{not} include the seed item, as we supply recommendations for cold items. Instead, they help us learn how categories are related. Since these labeled pairs can be sparse for certain categories, we furthermore incorporate ideas from Cycle Generative Adversarial Networks (CycleGAN)~\cite{zhu2017unpaired} in our architecture. This allows us to leverage available item information even when no labeled data exists for a given item and category.

Overall, we therefore suggest a novel \emph{semi-supervised} model for complementary item recommendations.
An evaluation of the model on three e-commerce datasets confirms that it is favorable over alternative approaches under different experimental settings. Moreover, the experiments clearly demonstrate the benefits of extending the architecture with CycleGAN elements. To our knowledge, utilizing CycleGAN concepts for improved complementary item recommendation has not been explored in the literature before.


\section{Background and Related Work}
\label{sec:related-work}

A common technical approach in real-world e-commerce is to determine related items, e.g., on item detail pages\footnote{Such recommendations are often not personalized according to long-term user profiles, which is also one assumption of our present work. See \cite{wang2021personalized} for a personalized approach at eBay.}, is to rely on co-purchasing patterns. Recommendations based on such simple shopping basket analyses 
can be surprisingly effective \cite{LeeHosanagar2018,Ludewig2018}. We note, however, that the items returned by such approaches are not necessarily complementary items, and it may happen that two particular shirts are frequently bought together \cite{Hao2020PCompanion}. In our work, we therefore assume 
that a complementary item belongs to one among the pre-defined complementary categories, which are different from the seed item's category (e.g., pants in the case of shirts).


Another type of recommendations commonly seen in practice are often shown under labels such as ``Similar Items'' or ``Related Items'', some of which may serve as substitutes \cite{McAuleyImage2015,Yao2018Judging,TrattnerJannach2019,Wang2011Utilizing,brovman2016optimizing}. Finding similar items is another relevant problem, but almost the opposite of our research focus. \emph{Related items}, see, e.g., \cite{Wang2011Utilizing}, on the other hand, can in theory include both complements (e.g., accessories) and substitutes, i.e., alternatives. Several works \cite{ahsan2020complementary,mane2019complementary,angelovska2021siamese} aim to find complementary products, but these often cannot distinguish between similar and complementary items, since the seed and the recommended items are encoded by the exact same network.


In terms of the addressed research challenge, the work by Galron et al.~\cite{galron2018deep} at eBay 
is closest to our research, and we also adopt a similar evaluation approach. Given the sparsity of user-item interaction data, which hampers the use of traditional collaborative filtering approaches, the item-based DCF (Deep Collaborative Filtering) method proposed in~\cite{galron2018deep} learns a similarity function between items based on their side-information. This similarity function is then used to retrieve recommendations for a given seed item, i.e., the neural network takes a pair of seed and target items as inputs, and returns a similarity prediction as an output. Internally, the network encodes the items as sparse feature vectors based on characteristics such as the title or category. To train the model, purchase data from eBay in the form of pairs of items that were co-purchased by the same users is used.
Computational experiments and an A/B test indicate that DCF performed significantly better than the existing system at eBay. 
Given the proven performance of DCF in a real-world setting, we use this method as a baseline in our research. We note that in the evaluation of DCF recommendations from the same category were considered as ground truth. 

\emph{P-Companion} \cite{Hao2020PCompanion} is another neural framework for complementary item recommendation. This model focuses on diversifying the recommendations and as it outperformed several other methods \cite{McAuley2015Inferring,Wang2018PathConstrained,hao2019universal}, we also use it as baseline. While this model has some commonalities with our approach, it does not allow semi-supervised learning. Furthermore, it does not maintain a separate latent space per category. Instead, the various dimensions in the item embeddings are weighted differently, according to their category.

In terms of applications, various previous works focus on fashion recommendation \cite{zhao2017deep,de2022disentangling,wu2019session,li2020learning}. In many cases, \emph{visual} approaches are highly effective \cite{cohen2021black}. Such approaches typically consider domain-specifics and strongly rely on item images to establish relationships between different clothing items, e.g., by projecting items in a shared visual style space \cite{McAuley2015Inferring}. In a recent work \cite{de2022disentangling}, the authors for example use shape and color information to learn which shapes and colors are compatible. Compatibility estimation is also the focus in \cite{Song2018Neural,He2016LearningComp}. A session-based recommendation approach, which also considers visual information is presented in \cite{wu2019session}. However, such visual approaches can also have their limitations. Specifically, purely visual approaches may return items that are stylistically similar, but not complementary. In \cite{zhao2017deep}, the authors therefore propose a complementary item recommendation approach that relies on textual information, and can be extended using recent methods \cite{shalom2021natural}. These works are different from our approach in that our framework is not domain-specific. Furthermore, our approach considers both visual and textual side-information and in addition allows us to specify target categories for the complementary items. Explicitly making compatibility predictions between items is not in the focus of our present work, but an interesting area for future investigations. Similarly, works that aim to automatically discern if two items rather represent alternatives or complements, e.g., \cite{LiuDecoupled2020,Wang2018PathConstrained}, may be integrated in our framework in the future as well to assess if they help further improve accuracy. 

A number of other works exist that use datasets that contain information about co-purchased or co-viewed items like we do in our computational experiments, e.g., \cite{Trofimov2018InferringComp,Zaho2017Improving}. The focus of such works however is often not primarily on making complementary item recommendations. Instead, the goal is to improve prediction accuracy in general, which is achieved by exploiting co-occurrence information in the data. No distinction is however made if the resulting recommendations actually contain substitutes or complements.

The use of adversarial training in recommender systems is widespread \cite{krishnan2018adversarial,krishnan2019modular,resheff2018privacy}. Specifically, using GANs for complementary recommendations was proposed in \cite{kumar2019c}, where a generator learns how to create the image of a complementary item. However,
the goal of this methods is to generate an image and not an item.
Another generative adversarial learning approach is proposed in \cite{huynh2018craft}. Given a seed item, a generator network creates recommendations and a discriminator needs to distinguish between real labeled recommendations and generated ones. However, this model requires full supervision and, unlike our approach, it does not allow to control important traits of the recommendations, like the category.

A \emph{quality-aware} method for complementary item recommendation is proposed in \cite{zhang2018quality}. The underlying intuition is that not only the compatibility of the items matters, but also the quality of the recommended items. The quality in their approach is based on explicit item ratings, and the proposed model jointly considers user preferences and compatibility aspects. In our work, in contrast, we do not assume the existence of long-term user preference profiles and explicit ratings. Nonetheless, personalizing the recommendations \cite{yan2022personalized, wang2022learning, zhang2021learning} or considering the sequential nature of user-sessions \cite{xu2020product} may be an aspect to explore in the future.



\begin{figure*}
\centering
\includegraphics[width=0.9\linewidth]{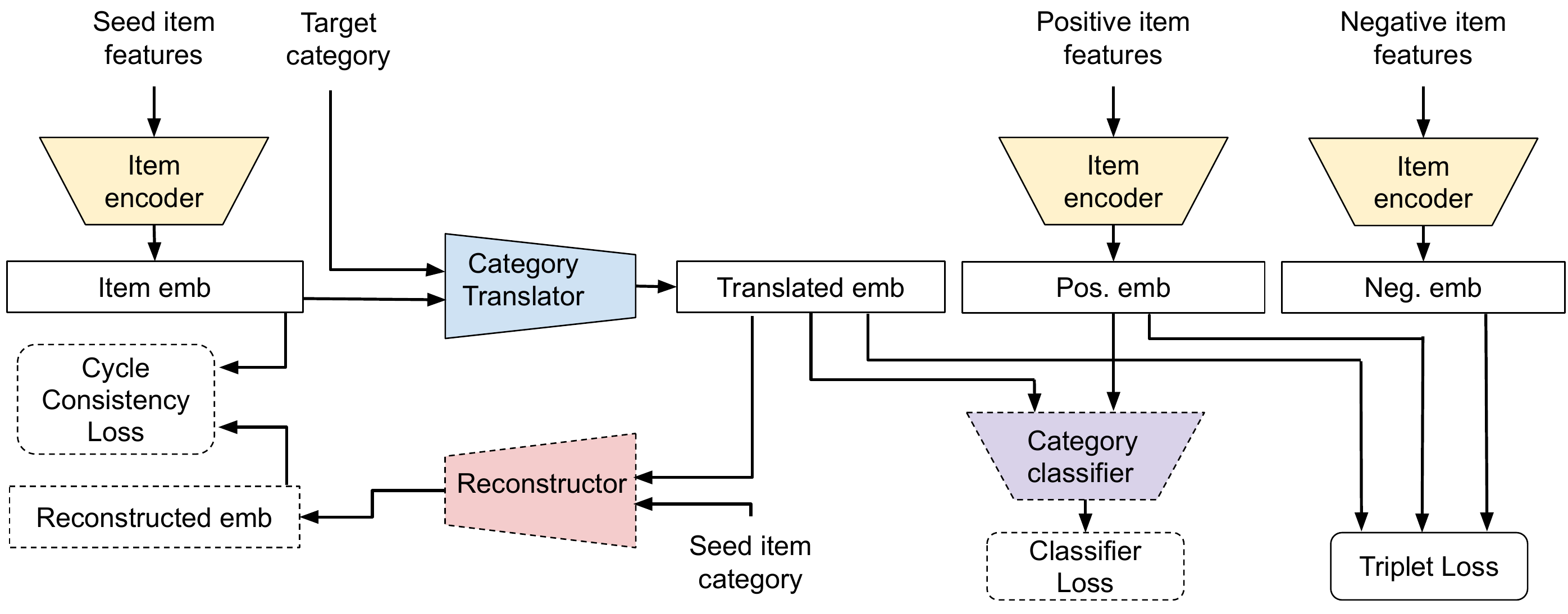}
\caption{The proposed model architecture. Elements with dashed lines are part of the adversarial learning network.}
\label{fig:arch}
\end{figure*}

\section{Overview of Proposed Approach}\label{sec:approach}
\paragraph{Summary of Problem Setting}
We recall the specifics of our problem setting. Given a \emph{seed} item and a \emph{target} category, the task is to recommend complementary items from the target category\footnote{In practice, there can be multiple complementary categories and the recommendation process may thus be repeated for each category.
}. The main challenge is that no interaction data yet exists for the seed item. However, we assume that each item has a known category and there is some additional side-information available for each item. Moreover, we make the assumption that there exists a labeled set of pairs of complementary items\footnote{Note that the pairs in the labeled set do not contain the seed item.}. Such a set may be created manually or automatically derived, e.g.,
by analyzing co-purchase patterns of pairs of items with sufficient purchase signals.
Generally, however, paucity of labeled data is an inherent problem, and a model that can cope with it is preferable.




\paragraph{Technical Approach}
\newcommand{\Triplets}{Supervised Translation\xspace}
\newcommand{\CycleGAN}{Adversarial Translation\xspace}

Let $\mathcal{C}$ denote the set of all item categories in the catalog.
As mentioned, one key idea of our approach is to maintain a latent space for each item category $c \in C$. To make a recommendation for a given seed item, we first create a distributed representation (i.e., embedding) of it, and we then translate this embedding into the target category's latent space. The translation process can be understood as creating a pseudo-item by converting the seed item's representation into the latent space of the target category. The main challenge in our approach now is to learn based on sparse data how to translate items to the target latent space in a way that the important traits of the seed item are maintained. Once the seed item is positioned in the target category's latent space, plausible recommendations can be derived by selecting items that are close to it. More formally, let $I_c$ be the set of items in category $c$. To find plausible recommendations in category $c$ for seed item $s$, the translated representation $v^c_s$ is generated. Then items are recommended according to their distance from $v^c_s$:
$\operatorname{argmax}_{i\in I_c} \operatorname{cos}(v^c_s, v_i)$, where $\operatorname{cos}$ is the cosine similarity and $v_i$ is a distributed representation of item $i$.

We recall that in our approach we aim to concentrate on a subset of \emph{relevant} categories for each item. Therefore, given an item's category, we determine complementary categories from item-level labeled data. To find the set of complementary categories for category $c \in \mathcal{C}$ we aggregate the item-level labeled pairs to the category-level to determine the set of other relevant categories\footnote{In our work we do not allow complementary item recommendations from the same category, as done in \cite{wu2019session}. This constraint can however be trivially dismissed as needed.}

\paragraph{Architecture Components}
The overall architecture has two main elements. First, the architecture includes a \emph{supervised} learning component, which uses item side-information and the labeled set of pairs of complementary items to learn the translation between the category latent spaces. Since this labeled set of pairs may suffer from data paucity in particular for rare categories, we extend the architecture to include a sub-network that supports \emph{unsupervised} adversarial learning from all items in the catalog. I.e., also for items for which there is no labeled complementary item in a given category in the dataset. 
Thus, the two elements combined together constitute a semi-supervised learning approach. 
Figure \ref{fig:arch} shows the overall architecture of the model.

\section{Supervised Learning Component}
\label{sec:triplets-recommender}
The supervised learning component has two elements. The \emph{item encoder} creates an embedding of the items.
 The \emph{category translator} 
then translates the encoded item into the latent space of any given target category.

\paragraph{Item encoder}
Given an item $i$, the model first generates its representation $v_i$ based on its side-information. In many application scenarios, including fashion as discussed above, the image of the item is a highly important piece of side-information. To incorporate this information, we first feed the item image through a pre-trained image processing model. Then, we pass its output through a learned multilayer perceptron (MLP). For categorical features, the model fits an embedding per each distinct value. We note that continuous features, such as an item's price, can be dealt with by converting them to categorical features through discretization. To obtain the overall item representation, the representations of all features are concatenated and passed through another MLP.
We acknowledge that other, more sophisticated approaches 
could be used instead of a simple MLP. Yet, we defer these potential improvements to future work.
We note that the category itself is used here as another feature, which makes the \emph{item encoder} category-aware.

\paragraph{Category translator} The task of this element is to learn how to transfer the item representations from one latent space to another. For example, let $s$ be a specific shirt and $p$ and $n$ (stand for positive and negative) are two pairs of pants, with embeddings $v_s$, $v_p$ and $v_n$, respectively. Let us assume that $s$ is complemented by $p$, but not by $n$. Then, we would like to learn a transformation such that projecting the shirt's representation $v_s$ into the pants domain will yield a representation that is more similar to $v_p$ than to $v_n$. We stress that there are many plausible architectures to combine the item and the category representations, some of them allow heterogeneous dimensionalities of the latent spaces of the various categories. However, our model of choice is a simple concatenation of these embeddings, followed by an MLP.

\paragraph{Model training} Training in the supervised learning component is based on the given set of labeled pairs of items. Each labeled pair consist of a seed item $s$ and a positive complementary item $p$ of category $c$. For each such pair, we apply negative sampling and draw a negative item $n$ (from category $c$ as well).
The \emph{item encoder} is then invoked on each of the items to obtain the representations $v_s$, $v_p$ and $v_n$. Then, applying the \emph{category translator} on $v_s$ yields $v^c_s$, the representation of the seed item in the desired latent space. As a loss function, we use the triplet loss~\cite{schroff2015facenet}: $\mathcal {L}\left(v^c_s,v_p,v_n\right)=\operatorname {max} \left({\operatorname {d} \left(v^c_s,v_p\right)}-{\operatorname {d} \left(v^c_s,v_n\right)}+\alpha ,0\right)$, where $\operatorname {d}$ is a distance function; the negative cosine similarity performed well in our experiments. $\alpha$ is a hyperparameter that sets the margin by which the positive sample should be more similar than the negative one.

\section{Adversarial Learning Component}
\label{sec:cyclegan-recommender}
The effectiveness of any \emph{supervised} learning approach is bound by the available labeled data. In complementary item recommendation, data paucity, as mentioned, is a common problem. In particular for more rare categories there may not be a sufficient amount of labeled pairs available for effective learning. Therefore, our architecture includes an adversarial learning component which implements \emph{unsupervised} learning. Thereby, we aim to leverage information from any item in the catalog, even if there are no labeled complementary items for it in a given category. In our example, this would be the case if we are given a particular shirt and there is no complementary item in the pants category in the training data.
Ultimately, the combination of both components leads to a \emph{semi-supervised} learning approach.
We emphasize that the adversarial component is only needed to further improve the training process. At inference time, only the outputs of the \emph{item encoder} and the \emph{category translator} are used to determine the recommendations.

\subsection{Architecture Elements}
The main goal of this component is to further improve the effectiveness of the \emph{category translator}. To ensure that the \emph{category translator} is not misguided when incorporating information about items without labels, the adversarial component takes inspiration from CycleGAN~\cite{zhu2017unpaired}. CycleGAN was designed in the context image-to-image translation problems, e.g., to translate a summer landscape into winter. A central idea in these networks is the concept of \emph{cycle consistency}. For the mentioned example, a translation would be cycle consistent if we end up close to the original image if we translate the winter landscape back. Ultimately, cycle consistency ensures that inputs and outputs are paired in a meaningful way.

Technically, our adversarial learning component has two main sub-networks, the \emph{classifier} and the \emph{reconstructor}.
The \emph{classifier} 
receives an item representation $v_i$ and a category $c$ as an input and returns the probability that item $i$ belongs to category $c$. To this end, the \emph{classifier} assigns a score for each category by passing $v_i$ to an MLP, where the last layer is of size $|\mathcal{C}|$. Then, these scores are turned into probabilities using \emph{softmax} and the probability of the item belonging to category $c$ is extracted.
Finally, the translated representation $v^c_i$ is sent to the \emph{classifier} to affirm it indeed looks like a representation of an item in category $c$. This auxiliary component motivates the \emph{category translator} to produce reliable embeddings. However, it might be incapable to train a good recommender, because it is not guaranteed that important traits of the seed item are preserved. In our running example, the translated representation may seem like it belongs to a pair of pants, but it might not lead to good recommendations, as it is not constrained to retain either concrete or abstract properties of the original shirt like style, age group, or price.

To address this issue, the \emph{reconstructor} is introduced, which encourages the \emph{category translator} to be \emph{cycle consistent} and 
to retain the traits of the seed item. Specifically, this network receives a translated representation $v^c_i$ and the original category of the seed item. It then aims to reconstruct the original representation of the seed item $v_i$ by returning a vector $v'_i$ such that $v_i \approx v'_i$.

\subsection{Model training}
The adversarial learning component incorporates two types of loss functions: for the \emph{reconstructor} and for the \emph{classifier}. To allow end-to-end learning of the entire architecture, the final loss is given by a weighted sum of these two losses and the loss function of the supervised learning component. The weights for the losses constitute a convex combination and are determined by hyperparameters.

\paragraph{Cycle consistency loss} The loss of the \emph{reconstructor}, in our approach is the Euclidean distance $\|v_i-v'_i\|^2$.
Since both terms in the loss, $v_i$ and $v'_i$, are outputs of learned networks, a degenerated, yet optimal loss can be achieved. For instance, if the \emph{item encoder} and the \emph{category translator} return the same fixed value, regardless of the input, then the loss would be 0. To overcome this issue, we recall that the purpose of this loss is to motivate the \emph{category translator} to not dismiss important traits in the seed item, represented by $v_i$. Therefore, $v_i$ serves as a label for this loss. Consequently, we stop the gradients backpropagate from $v_i$ due to this loss.
We point out that this procedure allows to further improve the \emph{item encoder}, due to the Jacobian obtained from $v'_i$.

\paragraph{Classifier loss}
As mentioned above, given an item representation and a category, the \emph{classifier} outputs the probability $p$ that the represented item belongs to this category. Conventionally, its loss is cross-entropy and since it is supplied with a single true category, the loss is contracted to $-\!\operatorname{log} p$.
We should bear in mind that only a well trained classifier can challenge the \emph{category translator} and thereby allow it to generate suitable representations. This poses several difficulties and rules out the feasibility of a naive implementation of the architecture. We discuss these challenges and how we addressed them as follows.
To optimally train the \emph{classifier} we supply it with two types of training instances.

The first type consists of genuine outputs of the \emph{item encoder}. Namely, we invoke the \emph{classifier} twice, with the embeddings of the seed item $v_s$ and the positive item $v_p$\footnote{For running time considerations the negative item's embedding is not fed to the \emph{classifier}, although it could be trivially added.}, where the label categories are the ones from the catalog. However, we note that a standard implementation would also affect the \emph{item encoder}. Specifically, it can ultimately lead to a degenerated model, as the category is an input to the \emph{item encoder}.
That way, the \emph{item encoder} is motivated to cooperate with the \emph{category translator} so as to excel at this loss at the expense of the true objective of the model, which is supplying recommendations. For example, the \emph{item encoder} may output the category embedding, while ignoring the rest of its input features.
We therefore implemented the model in a way that we prevent the gradients of the \emph{classifier} of flowing through the parameters of the \emph{item encoder}. Our experiments revealed that this led to significantly improved performance.

As for the second type of input for the \emph{classifier}, we recall that the \emph{category translator} aims to create an 
embedding $v^c_s$, which makes the \emph{classifier} classify $v^c_s$ as if it belongs to category $c$. In order to avert a situation where the \emph{category translator} finds edge cases (adversarial examples) that only deceive the \emph{classifier}, but do not look like real vectors of items in category $c$, it is important that the \emph{classifier} is trained on the output of the \emph{category translator} $v^c_s$, with label category $c$. If done this way, $v^c_s$ is used to train both the \emph{category translator} and the \emph{classifier}, but each of them has a different objective. The \emph{category translator} aims to fool the \emph{classifier}, while the latter needs to challenge the \emph{category translator}. To address this issue, we use \emph{adversarial training} with a gradient reversal layer. This means that during backpropagation the \emph{category translator} obtains the original gradients, while the \emph{classifier} is directed by the additive inverse of the gradients.

We note that this model is scalable since its training time is linear in the size of the labeled set and item-category pairs; also, the amount of parameters grows linearly with the number of categories.

\subsection{Specific CycleGAN Adaptations}

Here, we lay out the commonalities and differences of CycleGAN and our approach in more detail. Specifically, this exposition will clarify how we transferred ideas from CycleGAN to the complementary item recommendation problem in an innovative way.

CycleGAN is a variant of Generative Adversarial Networks (GAN) \cite{goodfellow2020generative}, which includes two generators and two discriminators that are trained simultaneously.
Let $X$ be the source domain and $Y$ the target domain for a given translation problem (e.g., summer to winter). Generator \texttt{G} learns to transform images from $X$ to $Y$. Generator \texttt{F} learns to transform images from $Y$ to $X$.
Discriminator \texttt{D\_X} learns to differentiate between genuine images of $X$ and generated images~$F(Y)$. That is, its objective is to return a high probability value for $x\in X$ and a low probability value for $\texttt{F}(y)$, for $y\in Y$. Similarly, discriminator \texttt{D\_Y} operates in domain $Y$. For simplicity, we collectively refer to \texttt{D\_X} and \texttt{D\_Y} as \texttt{D}, which represents a discriminator that distinguishes between genuine and generated images.
Training of these networks is done using adversarial training. It is generally desired that $\texttt{G}(x)$ does not dismiss important characteristics of the original image. This would guarantee that $\texttt{G}(x)$ is a translation of $x$, rather than just an image that seems to belong to $Y$, but has no affinity to $x$. To this end, the cycle consistency loss is introduced. Namely, this loss asks to minimize the difference between $x$ and $\texttt{F}(\texttt{G}(x))$.

We first explain that in a sense, the \emph{item encoder} serves as the real distribution, the \emph{category translator} as \texttt{G}, the \emph{classifier} as \texttt{D}, and the \emph{reconstructor} as \texttt{F}. Given an input representation $v$ and a category $c$, the \emph{classifier} works as follows. If $v$ is a was generated directly by the \emph{item encoder}, then it should confirm that the item belongs to the declared category $c$; otherwise, $v$ was further translated by the \emph{category translator}, and the \emph{classifier} should reject an assumed affinity between the item and the category. Therefore, similar to GANs, the \emph{classifier} (\texttt{D}) aims to output a probability close to one for inputs drawn from the \emph{item encoder} (real distribution) and probability of zero for inputs from the \emph{category translator} (\texttt{G}).
Like in CycleGAN, our approach applies the cycle consistency loss using the \emph{reconstructor}, which translates the item representation back to its original category and therefore serves as \texttt{F}.

However, there are some notable differences between the two models. First, the instances drawn from the real distribution in CycleGAN are genuine images from the source domain. Therefore, the cycle consistency loss is well defined, as the original image $x$ serves as the label for $\texttt{F}(\texttt{G}(x))$. In contrast, our approach works on the feature space. That is, the ``real distribution'' is actually the output of the \emph{item encoder}, which is a \emph{learned} network. Consequently, the outputs of the \emph{image encoder} also serve as labels, which might hamper its training. As mentioned before, we overcome this problem by stopping the gradients that arise from the seed item's vector.

Another difference stems from the complexity of the \emph{category translator}. While in CycleGAN there are only two domains, in our problem the number of domains (categories) can reach to hundreds. Therefore we resort to training a unified translator for all categories. Given the unified translator it might seem like the \emph{reconstructor} is redundant, as also the \emph{category translator} can translate between any pair of categories. However, in our experiments we noticed a crucial advantage in separating these two networks.
We conjecture that it is due to their different objectives. The \emph{category translator} aims to find good recommendations, while the \emph{reconstructor} wishes to recover the original latent traits of the item. Therefore, the \emph{category translator} may generate vectors that are similar to those of popular or high quality items, since they usually make good recommendations, while the \emph{reconstructor} is not bound to this need.
Therefore, the \emph{category translator}, which generates the recommendations, should preferably be distinct from the \emph{reconstructor}.

\section{Experimental Evaluation}\label{sec:experiments}
We conducted an in-depth experimental evaluation of our approach. All our code is shared online for reproducibility
\footnote{\url{https://fb.me/cgan_ccomplementary_item_recommendation}}.

\subsection{Experiment Design}
\paragraph{Datasets \& Preprocessing}
We rely on real-world data from \emph{Amazon} \cite{he2016ups}, as used in previous related works~\cite{LiuDecoupled2020,Hao2020PCompanion}.
From these datasets, we first extracted the items' side-information, which include image, price, and category. Furthermore, each item $i$ in the dataset can be associated with a list of recommended items $\{r_i\}$ that are \emph{frequently bought together}. 
By removing items from $\{r_i\}$ that belong to the same category as $i$, we create a labeled set of complementary item pairs $\{(s, r_i)\}$, which we use for model training and evaluation in the experiments. We note that these pairs may be asymmetric, e.g., a laptop may be accompanied by a recommendation for a charger, but not vice versa.

We considered three subsets of the \emph{Amazon} datasets: ``Clothing, Shoes and Jewelry'' (dubbed as Clothing), ``Home and Kitchen'' (Home), and ``Toys and Games'' (Toys). These subsets were chosen because complementary items are of key importance in these domains. At the same time, they differ in key aspects like their verticals, target users, attributes that affect the notion of complementary, and number of items and categories. The main statistics of these datasets are shown in Table~\ref{tab:dataset_statistics}.

\begin{table}[tbh]
    \centering
    \caption{Dataset statistics}
    \begin{tabular}{lccc}
\toprule
Statistic                   & Clothing & Toys & Home\\
\midrule
\#items                     & 14,591 & 20,510 & 29,258 \\
\#item pairs                & 53,375 & 112,964    & 162,497 \\
\#categories                & 126   & 124       & 286 \\ 
\#category pairs            & 4,621  & 5,734      & 18,999 \\ 
Avg. items per category     & 115.8 & 165.4     & 102.3 \\ 
Max items per category      & 808   & 2,558      & 796 \\ 
Min items per category      & 12    & 16        & 11 \\ 
\bottomrule
\end{tabular}
    \label{tab:dataset_statistics}
\end{table}

In terms of pre-processing, we excluded categories with less than five items to reduce noise. As a pre-trained image processing model, we used ResNet152~\cite{he2016deep}. If an item had multiple images, we consider only the first one. We furthermore discretized the continuous price to twenty bins using equal-depth binning. We used a random sample of 80\% of the data for training and the rest for validation and testing. 
To mimic the cold start problem, we ensured that the seed items in the validation and test sets do not appear in the training set.

\paragraph{Evaluation Procedure \& Metrics}
The output of our model is a ranked list of complementary items, given a seed item. We therefore apply standard list-based accuracy metrics, namely Hit Rate ($HR@k$) and NDCG (Normalized Discounted Cumulative Gain), see \cite{shani2011evaluating}. Moreover, since the labeled set is highly skewed to popular items, some models may focus on a narrow set of such popular items in their recommendations, thus leading to limited coverage and diversity. Therefore, we also report \emph{catalog coverage}~\cite{Herlocker:2004:ECF:963770.963772}, which is defined as the fraction of the items of the catalog that appear in the top-$k$ recommendation lists for the seed items in the test set. The value of $k$ in this measure was 10.

Following the described problem setting, in the main evaluation protocol the desired category of the recommendations is given. 
To showcase the robustness of the methods, we also experiment in a setting where the recommendations can come from any category. We recall that our model requires a target category as an input. To create a recommendation list that considers all relevant categories, we go over all complementary categories of the seed item in a round-robin fashion and iteratively select the next recommendation.

\paragraph{Baselines}
We compare our model against these baselines.

\begin{itemize}
\item \textbf{Popularity} is a simple yet effective baseline, which utilizes the labeled set to count the popularity of each item. For each item $i$ it records the number of seed items for which $i$ is labeled as their complementary item in the labeled set.
\item \textbf{DCF}~\cite{galron2018deep} is a recent neural model optimized for cold items. Since its code was not released, we implemented it based on the 
    the original paper and make it publicly available.
\item \textbf{DCF-Hard} is a variant of \emph{DCF} that we propose here. It leverages category information to apply \emph{hard negative mining}. Specifically, negative samples are not drawn from the entire catalog, but only from the items in the category of the labeled target item.
\item \textbf{P-Companion}~\cite{Hao2020PCompanion} is another recent neural model. Since its code is not publicly available, we implemented it and publish the code in our repository as well. To make it compatible with our problem setting we made two modifications. First, we control the target categories and second, we omit collaborative information to support the cold start problem.
\end{itemize}

We carefully tuned the hyperparameters both of our model and of the baselines for each of the datasets in a manual process. The final hyperparameters can be found in the code repository.

\subsection{Results}

\subsubsection{Main Results}
Table \ref{tab:catagory_aware_fbt_perf} reports the main results of our experiments, where the category is given to the recommenders. We name our proposed model \textbf{ALCIR}, which stands for \emph{\textbf{A}dversarial \textbf{L}earning for \textbf{C}omplementary \textbf{I}tem \textbf{R}ecommendation}. To analyze the contribution of the individual elements of our architecture, we report results for (a) \emph{ALCIR-Sup}, which only consists of the supervised learning component from Section~\ref{sec:triplets-recommender} and (b) \emph{ALCIR}, which corresponds to the \emph{full} model as described in Section \ref{sec:cyclegan-recommender}.

\begin{table}[htb]
\caption{Category-aware recommendation performance}
\label{tab:catagory_aware_fbt_perf}
\begin{tabular}{lccccc} 
\toprule
Method                   & NDCG                      & HR@1            & HR@5           & HR@10  & Cov. (\%)         \\ 
\midrule
\rowcolor{lightgray}\multicolumn{6}{l}{\textbf{Clothing Dataset}}\\
Popularity            & 0.238                     & 0.051                  & 0.153          & 0.234 &   6.99       \\
                             DCF                         & 0.211                     & 0.011                  & 0.059          & 0.111  &   7.24      \\
                             DCF-Hard & 0.226                     & 0.019                   & 0.083          & 0.147   &  40.49      \\
                             P-Companion             &  0.238	                & 0.030	                   & 0.102	            & 0.169	    & 84.59     \\
                             ALCIR-Sup          & 0.298	                    & 0.074	        	         & 0.203	       & 0.302  & \textbf{89.92} \\
                             ALCIR         &  \textbf{0.316}            & \textbf{0.092}  & \textbf{0.233} & \textbf{0.332} & 88.5  \\ 
\midrule
\rowcolor{lightgray}\multicolumn{6}{l}{\textbf{Toys Dataset}}\\
Popularity  & 0.231                     & 0.038                   & 0.126          & 0.200  &   5.13      \\
                             DCF                     & 0.193                     & 0.009                    & 0.042          & 0.081  &   5.96      \\
                             DCF-Hard &  0.208                     & 0.017           & 0.064          & 0.111  &    36.72     \\
                             P-Companion             &  0.235	                   & 0.028	         	      & 0.104	    & 0.172	    & \textbf{93.78}       \\
                             ALCIR-Sup         & 0.297                     & 0.073                    & 0.205          & 0.303  &   90.46      \\
                             ALCIR         &  \textbf{0.308}            & \textbf{0.078}  & \textbf{0.224} & \textbf{0.330} & 82.62 \\ 
\midrule
\rowcolor{lightgray}\multicolumn{6}{l}{\textbf{Home Dataset}}\\
Popularity   & 0.251                     & 0.048                    & 0.160          & 0.255  &    8.2     \\
                             DCF             & 0.211                     & 0.011                    & 0.051          & 0.102  &   8.32      \\
                             DCF-Hard         & 0.221                             & 0.014          & 0.046          & 0.077  &   12.82      \\
                             P-Companion           	& 0.245	& 0.032		& 0.111	& 0.187	& \textbf{94.45}     \\
                             ALCIR-Sup        & 0.296                     & 0.068                   & 0.197          & 0.293  &    93.44     \\
                             ALCIR         &  \textbf{0.304}            & \textbf{0.077}  & \textbf{0.210} & \textbf{0.312} & 94.38  \\
\bottomrule
\end{tabular}

\end{table}

The results in Table \ref{tab:catagory_aware_fbt_perf} show that \emph{ALCIR} consistently outperforms all baselines in terms of the accuracy measures on all three datasets, usually with a large margin\footnote{We report results at additional cut-off thresholds in the code repository.}. We additionally observe that already the supervised component of our model (\emph{ALCIR-Supervised}) performs better than the baselines. The adversarial component is then successful in even further increasing these already strong results.

Considering the ranking of the other models, we notice that the popularity-based approach represents a baseline that can be difficult to beat. The \emph{DCF} model and even the improved \emph{DCF-Hard} model never reach the accuracy levels of the popularity-based method. \emph{P-Companion} works better, but still does not reach the Hit Rate values of popular-item recommendations. Only in terms of the NDCG, \emph{P-Companion} reaches a similar performance level. The strong performance of the popularity-based method is not surprising, though. An inspection of the datasets revealed that 
the ten most popular items cover between a fifth and a quarter of the labeled complementary items. Moreover, also the evaluation of a  ``Co-Purchase'' method in \cite{Hao2020PCompanion} showed that \emph{P-Companion} did not outperform such a popularity-based approach in a user-centric study\footnote{Popularity-based recommendations were not examined in \cite{galron2018deep}.}.

Our proposed model, in contrast, outperforms the popularity-based model consistently. We also observe that \emph{ALCIR} leads to high \emph{coverage} values, ranging from 82\% to 94\%. Only the \emph{P-Companion} method reaches an even slightly higher catalog coverage. The popularity-based method by design only recommends very popular items, leading to the lowest coverage. Interestingly, also the \emph{DCF} method has very low catalog coverage. The improved DCF version (\emph{DCF-Hard}) helps to address the coverage problem to a good extent.

Table \ref{tab:perf_category_unaware} finally shows the performance results when we consider all relevant categories for complementary item recommendations as described above. Naturally, across all models, the performance results for this experiment are lower compared to a case where the category is known. Nevertheless, we observe that the proposed model ALCIR is superior also in this problem setting.

\begin{table}[h!tb]
\caption{Recommendation performance w/o target category}
\label{tab:perf_category_unaware}
\begin{tabular}{lccccc} 
\toprule
Method                  & NDCG           & HR@1                     & HR@5           & HR@10 & Cov. (\%)          \\ 
\midrule
 \rowcolor{lightgray}
\multicolumn{6}{l}{\textbf{Clothing Dataset}}\\
Popularity     & 0.099          & 0.001                   & 0.005          & 0.007 &  0.06          \\
                          DCF               & 0.111          & 0.000                   & 0.001          & 0.002 & 0.09          \\
                          DCF-Hard   & 0.118          & 0.001                    & 0.003          & 0.007 & 1.82          \\
                          P-Companion                      & 0.111          & 0.000                  & 0.002          & 0.003  & 0.62         \\
                          ALCIR-Sup                  & 0.150          & 0.009                   & 0.036          & 0.060 &  59.36        \\
                          ALCIR        &  \textbf{0.170} & \textbf{0.012}  & \textbf{0.060} & \textbf{0.098} & \textbf{60.55}  \\ 
\midrule
\rowcolor{lightgray}\multicolumn{6}{l}{\textbf{Toys Dataset}}\\
Popularity     & 0.102          & 0.001                    & 0.003          & 0.005 &  0.04         \\
                          DCF                  & 0.124          & 0.000                   & 0.001          & 0.002 & 0.06          \\
                          DCF-Hard   & 0.148          & 0.003                  & 0.010          & 0.016 &  5.67        \\
                          P-Companion                      & 0.123	                 & 0.001	          & 0.001	       & 0.002	 & 5.12         \\
                          ALCIR-Sup                    & 0.172          & \textbf{0.010}           & 0.044          & 0.021 &  \textbf{64.56}        \\
                          ALCIR         &  \textbf{0.184} & 0.009       & \textbf{0.046} & \textbf{0.071} & 56.03 \\ 
\midrule
\rowcolor{lightgray}\multicolumn{6}{l}{\textbf{Home Dataset}}\\
Popularity     & 0.093          & 0.000          & 0.001                & 0.002  & 0.03         \\
                          DCF                      & 0.117          & 0.000                   & 0.001          & 0.002  & 0.05         \\
                          DCF-Hard       & 0.117          & 0.000                  & 0.002          & 0.003  & 0.13         \\
                          P-Companion             	          & 0.117	       & 0.000	        	          & 0.001	      & 0.001	& 1.61         \\
                          ALCIR-Sup                     & 0.143          & 0.005                 & 0.021          & 0.033  & 53.42         \\
                          ALCIR         &  \textbf{0.150} & \textbf{0.008} & \textbf{0.030} & \textbf{0.049} &  \textbf{55.53} \\
\bottomrule
\end{tabular}

\end{table}

\begin{figure*}[h!tb]
\centering
\includegraphics[width=1.0\linewidth]{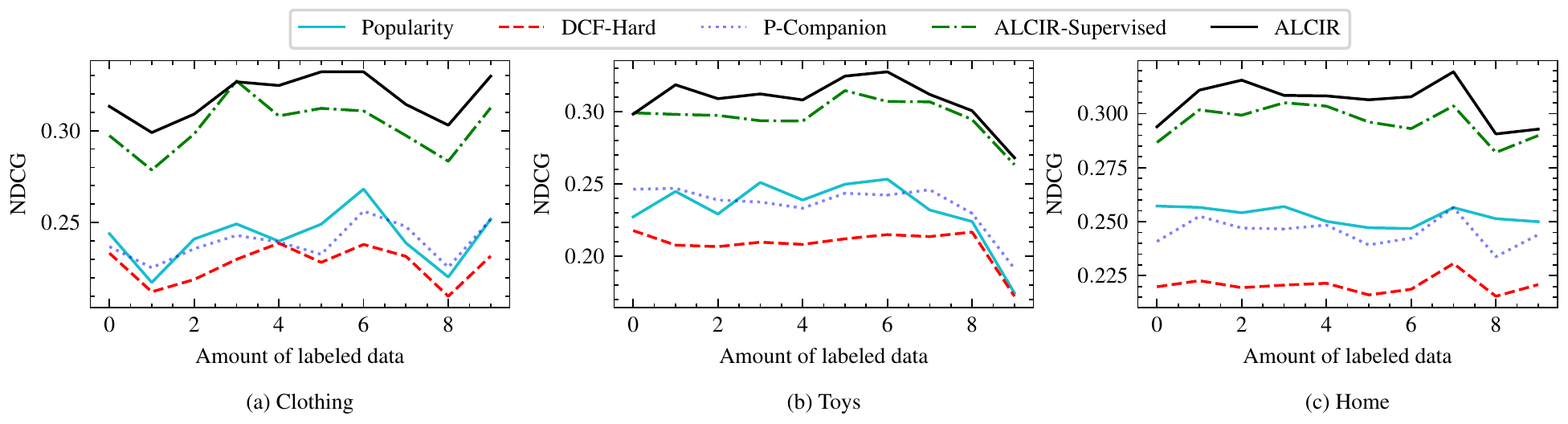}
\caption{Performance with respect to the amount of labeled data}
\label{fig:per_labeled_amount}
\end{figure*}

\subsubsection{Additional Analyses}
\paragraph{Impact of Data Paucity}
One main assumption when we introduced the unsupervised learning component to our model was that it will be particularly beneficial for rare categories. To validate this assumption, we conducted the following analysis to assess the relative contribution of the full \emph{ALCIR} method when we vary the amount of labeled data for each pair of categories. To this end, we count the number of labeled instances per each pair of complementary categories and discretize these counts to ten equally-sized bins. We order the bins by the amount of labeled data for each pair of categories in ascending order. The first bins therefore represent pairs of complementary categories for which little supervision is available in the training set.

Figure \ref{fig:per_labeled_amount} shows the performance of each model for the ten bins. We note that performance measures between different bins cannot be trivially compared, because each bin holds different target categories, with a different amount of items from which the recommender should select. Therefore, only the performance of the different models within the same bin and dataset can be compared, since they refer to the exact same test items. Considering this aspect, we can observe the same trend in all datasets, where the relative advantage of \emph{ALCIR} is higher for the rare category pairs, i.e, in the left parts of the figures. In contrast, the advantage of the full model becomes smaller for the very popular category pairs, and it is particularly pronounced for the Toys and Home datasets. Overall, the analysis confirms our assumption regarding the usefulness of the full model in particular for rare categories.

\paragraph{Ablation Study}
Besides consisting of a supervised and an unsupervised component, one central feature of our architecture is that it makes use of a number of specific loss functions (triplet loss, cycle consistency loss and classifier loss), all of which are aimed to increase the performance of the model. To validate that these architecture elements contribute to the overall model performance, we ran an ablation study to assess the performance when only some of the components are utilized. Table \ref{tab:ablation} shows the outcomes of this analysis in absolute numbers and relative to the performance of the complete model (shown in parentheses). The results provide evidence that indeed all components contribute to the performance of the model. Most noteworthy is that a model that does not utilize labeled data, but only the unsupervised losses of the classifier and the cycle consistency performs the worst. This, however, comes as no surprise, given the well-known importance of labeled data.

\begin{table}[htb]
\caption{Ablation study for category aware recommendation}
\label{tab:ablation}

\begin{tabular}{lp{1cm}p{1cm}p{1cm}p{1cm}} 
\toprule
\multicolumn{1}{c}{Method} & NDCG            & HR@1                  & HR@5            & HR@10            \\ 
\midrule
\rowcolor{lightgray}\multicolumn{5}{l}{\textbf{Clothing Dataset}}\\
 Classifier+Cycle          & 0.21   \newline (-33.5\%)  & 0.015 \newline (-83.9\%) & 0.059  \newline (-74.7\%) & 0.114  \newline  (-65.7\%)  \\
ALCIR-Sup                  & 0.298  \newline (-5.7\%)   & 0.074 \newline (-19.4\%)   & 0.203  \newline (-13.1\%) & 0.302 \newline  (-8.9\%)   \\
 Triplet+Cycle             & 0.311  \newline (-1.8\%)   & 0.087 \newline (-5.6\%)   & 0.226  \newline (-3.1\%)  & 0.319  \newline  (-3.8\%)   \\
Triplet+Classifier         & 0.305  \newline  (-3.5\%)	& 0.082 \newline (-11.3\%)	& 0.213 \newline (-8.8\%)	& 0.306  \newline (-7.5\%) \\
ALCIR                  & 0.316           & 0.092                & 0.233           & 0.332            \\ 
\midrule
\rowcolor{lightgray}\multicolumn{5}{l}{\textbf{Toys Dataset}}\\
 Classifier+Cycle           & 0.207  \newline (-32.9\%) & 0.016  \newline (-79.5\%)  & 0.07  \newline (-68.7\%)  & 0.124  \newline (-62.5\%)  \\
 ALCIR-Sup                  & 0.297  \newline (-3.7\%)  & 0.073  \newline (-6.4\%)   & 0.205 \newline (-8.7\%)  & 0.303  \newline  (-8.2\%)   \\
Triplet+Cycle               & 0.294  \newline (-4.4\%)  & 0.072  \newline (-6.8\%)   & 0.205 \newline (-8.3\%)  & 0.303  \newline (-8.4\%)   \\
Triplet+Classifier          & 0.298  \newline (-3.3\%)	& 0.073 \newline (-6.6\%)	 & 0.209 \newline (-6.6\%)	 & 0.309 \newline (-6.5\%) \\
 ALCIR                       & 0.308                     & 0.078                  & 0.224           & 0.330            \\ 
\midrule
\rowcolor{lightgray}\multicolumn{5}{l}{\textbf{Home Dataset}}\\
Classifier+Cycle           & 0.218  \newline (-28.1\%) & 0.014  \newline (-81.2\%)  & 0.068  \newline (-67.5\%) & 0.127  \newline (-59.3\%)  \\
ALCIR-Sup                  & 0.296  \newline (-2.6\%)  & 0.068  \newline (-11.4\%)  & 0.197  \newline (-6.1\%)  & 0.293  \newline (-6.1\%)   \\
Triplet+Cycle              & 0.298  \newline (-2\%)    & 0.069  \newline (-9.9\%)   & 0.201  \newline (-4.1\%)  & 0.303  \newline (-2.9\%)   \\
Triplet+Classifier         & 0.297 \newline (-2.1\%)             &	0.069 \newline (-10.7\%) &	0.199 \newline (-4.9\%) &	0.297 \newline (-4.7\%)          \\
ALCIR                  & 0.304           & 0.077               & 0.210           & 0.312            \\
\bottomrule
\end{tabular}

\end{table}

\section{Future Work}\label{sec:discussion}
In this work we have proposed a novel semi-supervised approach for the highly relevant problem of complementary item recommendation, and an in-depth empirical evaluation clearly demonstrates the benefits of the approach. Our insights point to a number of future research directions. Our work focused on complementary item recommendations for cold items, and we assume that existing data (e.g., about co-purchases) can be used for the warm items. In future work, we plan to  extend our model to also support warm items, using the same framework. That way, the input features for the item encoder would include also collaborative data by applying e.g., the method suggested in \cite{bibas2021single}. In addition to such extensions and continuing related research in \cite{wang2021personalized,zhang2018quality}, other promising approaches to further improve the effectiveness of the model could lie in the \emph{personalization} of the complementary item recommendations and to consider aspects of  item \emph{quality}.
Finally, our problem setting can be thought as a special case of domain adaptation, where we transfer representations from one category to another. An interesting future work would be to extend our method to other recommendation scenarios, like context-aware recommendations, where we generate item representations in different contexts.

\balance
\bibliographystyle{ACM-Reference-Format}
\bibliography{sample-base}

\end{document}